\documentstyle[11pt,epsf]{article}
\begin{document}

\vspace*{2cm}
\begin{center}
{\Huge\bf
Interaction of the Past \\ of parallel universes}
\vspace*{1.5cm}

{\large\bf  Alexander K.\ Guts}
\vspace*{1cm}

{\normalsize

Department of Mathematics, Omsk State University \\
644077 Omsk-77 RUSSIA}

\vspace*{0.5cm}
E-mail: guts@univer.omsk.su  \\
\vspace*{0.5cm}
October 26, 1999\\
\vspace{.5in}
\end{center}
\begin{center}
ABSTRACT
\end{center}

We constructed a model of five-dimensional Lorentz manifold with
foliation of codimension 1 the leaves of which are four-dimensional
space-times. The Past of these space-times can interact in macroscopic
scale by means of large quantum fluctuations. Hence, it is possible 
that our Human History consists of "somebody else's" (alien) events.

\newpage

\setcounter{page}{1}

%%%%%%%%%%%%%%%%%%%%%%%%%%%%%%%%%%%%%%%%%%%%%%%%%%%%%%%%%%%%%%%%%%%%
\def\R{{{\rm I} \! {\rm R}}}
\def\F{{\cal F}}
\def\D{{\cal D}}
%%%%%%%%%%%%%%%%%%%%%%%%%%%%%%

In this article  the possibility of interaction of the Past (or Future)
in macroscopic scales of
space and time of two different universes is analysed. Each universe
is considered  as four-dimensional space-time $V^4$, moreover they
are imbedded in five-dimensional Lorentz manifold $V^5$,
which shall below name Hyperspace.
The space-time $V^4$ is absolute world of events.
Consequently, from formal standpoints any point-event of this
manifold $V^4$, regardless of that we refer  it to Past, Present or
Future of some observer, is {\it equally available to operate  with her}.
In other words, modern theory of space-time, rising to Minkowsky,
postulates absolute eternity of the World of events in the sense
that all events exist always. Hence,  it is  possible
interaction of Present with Past and Future as well as Past
can interact with Future. Question is only in that as this is realized.
The numerous articles about time machine show that our statement on
the interaction of Present with Past is not fantasy, but is subject of the
scientific study.
In articles \cite{1,2,3} we used theory of foliations for construction one
of the possible ways of travel to the Past (Future). Exactly this
theory seems to be useful for decision  of problem on
interactions "nearby" universes.

So, it is assumed that manifold $V^5$ has foliation $\F$ of codimension 1,
but our Universe is a leaf $F_0^4$ in him. Other leaves represent different
universes.

Consider five-dimensional manifold that is got by  multiplying
on $\R^3$ of axial section of foliation of Reb in the torus
$S^1\times D^2$ (\cite[468]{Fom}, refer to Pic.1).

\vspace*{0.4cm}

\epsfbox[-100 150 100 100]{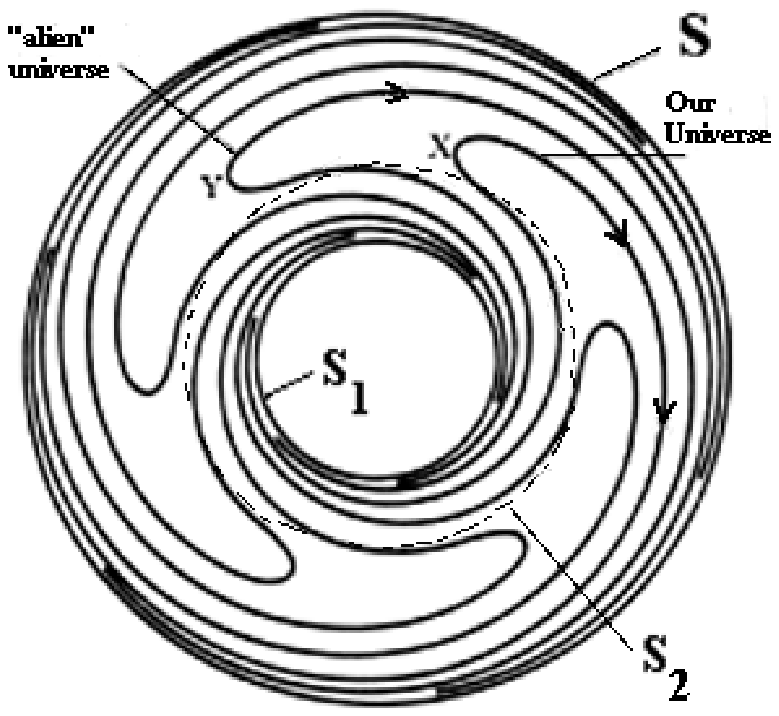}

\vspace*{7cm}
\begin{center}
{\small Pic.~1.
\vspace*{0cm}
\begin{center}
{\footnotesize Model Hyperspace of Reb  with interacting Past }
\end{center}
}
\end{center}

%\newpage

\newpage

Our Universe and we as its Observers are not single in this
mathematical  theory of Time. In Hyperspace other worlds are also situated.
Fix certain spatial section $X$ in leaf $F_0^4$. It represents "Present".
Similar "Present" we will fix in other leaves, for example, section $Y$
in leaf $F_1^4$ which is near to leaf $F_0^4$ (see Pic.1). We shall
consider only
those universes, i.e. leaves, which are situated in certain neighborhood
of our Universe and, accordingly, Present of "somebody else's" (alien)
universes
are situated in sufficiently small
neighborhood of "Present" of our Universe.

For this model of five-dimensional Reb Hyperspace   it is distinctive
that Past of our
Universe and Past of "someone else" (alien) universe
are approached that more strong,
than further from Present will run away past epoches. On Pic.~1 ring is
image of Hyperspace, curly lines are universes, each with
its Observers. One line  is our four-dimensional Universe $F_0^4$,
we are its Observers. Beside we see "someone else" (alien) four-dimensional
universe $F_1^4$ with its own observers.
Our Present is point $X$, Present of "someone else" universe is
a point $Y$. If we begin a trip  to
the Past against of the streem of time, we shall move in our World against
the arrow on line 1 (point $X$) and coil all more and more on the
circle $S_1$.  Similarly, travel in the Past in nearby universe
$F_1^4$ or line 2 (point $Y$) is a coiling motion around the circle $S_1$
along line 2 against the arrow of time.
Past of two worlds are approached in topology
of Hyperspace.
That can one occur?

For the answer to this question we will use geometrodynamics ideas
of Wheeler which we shall apply to five-dimensional theory of
gravitation.
Amplitude of probability of transition from Universe $F_0^4$
to universe $F_1^4$ will represent by means of
Feynman integral over 5-geometries:
$$
<F_0^4|F_1^4> = \int\limits_{F_0^4}^{F_1^4}\D g^{(5)}
 \exp \left[-\frac{iS}{\hbar}\right],\eqno(1)
$$
where
$$
S=\frac{c^3}{ 8\pi GT}\int R^{(5)}\sqrt {-g^{(5)}}d^5x \eqno(2)
$$
is action in five-dimensional Lorentz geometry \cite[p.52]{Vlad} with
metrics $g_{AB}^{(5)}$, moreover
$T$ is a constant with dimensionality [cm], connected with
5-th coordinate (for instance, it characterizes cyclicity on the
fifth coordinate in the Kaluza-Klein theories).
From (1), (2) it follows \cite{Wh} that (1) is not
changed under quantum fluctuations of five-dimensional
geometries $g_{AB}^{(5)}\ (A,B=1,...,5)$:
$$
\Delta g^{(5)} \sim \frac { L^*}{L}\sqrt{\frac{T}{L_0}},\eqno(3)
$$
where $L^*\sim 10^{-33} cm$ is constant of Plank, but $L^4 \times L_0$ is
a size of 5-region of fluctuations.

Formula (3) means that as soon as Past of universes $F_0^4$ and $F_1^4$2
are approached "sufficiently close", quantum fluctuations of metrics
begin to
change  topology and geometry of two universes; they begin to stick
together by  means of wormholes; one will appear the tunnel transition
between worlds.
This means that at least on the microscopic scale  the Past of these
two worlds are indistinguishable.

Formula (3) is not contradictory with classical four-dimensional formula
for quantum fluctuations
$$
\Delta g^{(4)} \sim \frac { L^*}{L},\eqno(4)
$$
because it was got under assumption $\sqrt{det \ g^{(4)}} \sim 1$ \cite{Mod}.

But fluctuations are significant on the macroscopic scale too.
In fact, suppose that $L \sim 1 \ km$. This corresponds
the time  interval $\sim 3\cdot 10^{-6} \ sec$.
Then, as it follows from (3), quantum fluctuation of 5-metrics
$\Delta g^{(5)} \sim 1$, if $L_0\sim 10^{-76}T$. In other words,
to begin the Past of our Universe and universe
$F_1^4$   to interact by means of
formation of wormholes between them in considered
model of Reb Hyperspace,  it
is necessary that it was sufficiently removed from Present.
Otherwise, to interact leaves $F_0^4$ and $F_1^4$ must
powerfully draw together.
Herewith  one interact spatial regions of size $ 1 km$, and time of
interaction is $10^{-6} \ sec$. For more extensive spatial regions
time of interactions increases. In principle it become
possible a transition between universes meaning  exchange of the Past.
Past our Universe can contain events which are not belonging to
our History.

Note that large quantum fluctuation, i.e. those that could arise at
large spatial scale, are essential detail of
four-dimensional quantum theory \cite{Mod}. In five-dimensional theory
one can be found an universe  $F$ which is contained
in  sufficiently thin neighborhood of our Universe. It follows from
(3) that there exist large quantum fluctuations which are the interactions
between Present of our Universe and "Present" of universe $F$.
The existence of such interactions is very  serious  question. It possible
that such interaction at scale $L_0< T_0, O<T_0<T$ in fifth dimension
are suppressed by, for example, scale-dependent cosmological term
$\Lambda(L,L_0)$ or some external field \cite{Mod}.

Hyperspace of Reb can be a subject of compression of part of
ring one border of which is cylinder $S_1$ and other is
cylinder $S_2$ labeled on Pic.1 by means of dotted line (Pic.1).
If $S_2$ tends to points $X,Y$, then interacting Past will all closer
to the current epoch.

Hyperspace of Reb can be a subject of local compression (Pic.2).
Then we shall have a model of periodic "strong" interactions of chosen
epoches of the Past.

\epsfbox[-100 120 30 30]{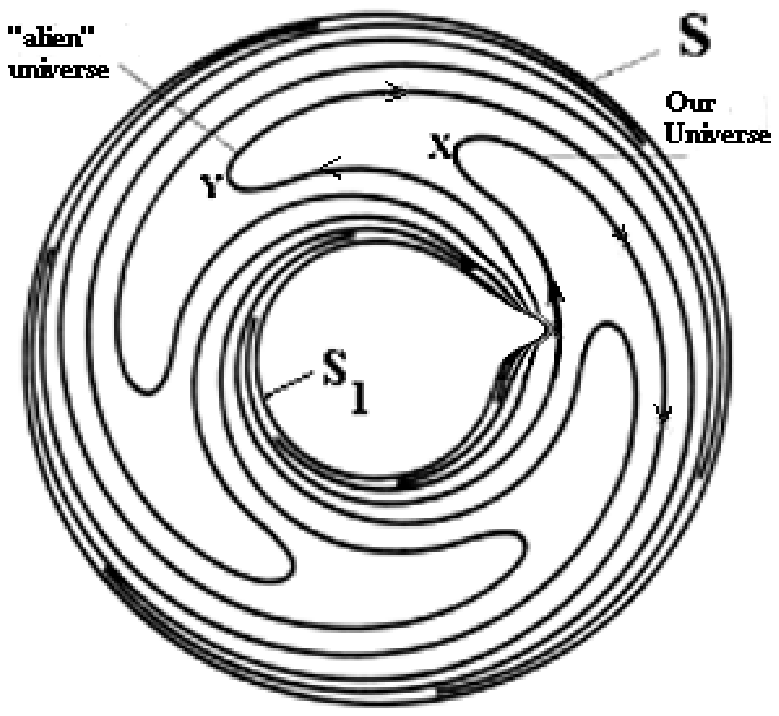}

\vspace*{7cm}
\begin{center}
{\small Pic.~2.
\vspace*{0cm}

%\begin{quote}
{\footnotesize  Model of hyperspace with
interacting nearby Past}
%\end{quote}

}
\end{center}

Wholly it can turn out to be that principal details of explored model
situation will found in the Reality and this has direct relations
to the problems in the historic science, which were open N.A.Morozov,
A.T.Fomenko and his co-authors  \cite{Imp}.
Historical text-books are  contradictory,  and this is objective
Law of Nature \cite{Gu}. Human History contains many different variants
of events. Maybe, one is openned
prospect of building of Multivariant World History of Human
Civilization which can  conciliate supporters of traditional and
new chronology \cite{Guc}.

\end{document}